**Electrical control of topological 3Q state in intercalated van der Waals antiferromagnet $Co_x$-$TaS_2$**

Junghyun Kim[1,2,3], Kai-Xuan Zhang[1,2,4]*, Pyeongjae Park[1,2,5], Woonghee Cho[1,2], Hyuncheol Kim[1,2], Han-Jin Noh[6], and Je-Geun Park[1,2,4]*

[1]Department of Physics and Astronomy, Seoul National University, Seoul 08826, South Korea
[2]Center for Quantum Materials, Department of Physics and Astronomy, Seoul National University, Seoul 08826, South Korea
[3]Center for Spintronics, Korea Institute of Science and Technology (KIST), Seoul 02792, South Korea
[4]Institute of Applied Physics, Seoul National University, Seoul 08826, South Korea
[5]Materials Science and Technology Division, Oak Ridge National Laboratory, Oak Ridge, Tennessee 37831, USA
[6]Department of Physics, Chonnam National University, Gwangju 61186, Republic of Korea

* Corresponding Authors: Kai-Xuan Zhang (kxzhang.research@gmail.com); Je-Geun Park (jgpark10@snu.ac.kr)

**Abstract**

Van der Waals (vdW) magnets have opened a new avenue of opportunities encompassing various interesting phases. $Co_{1/3}TaS_2$–an intercalated metallic vdW antiferromagnet–is one of the latest additions to this growing list of materials due to its unique topologically nontrivial triple-**Q** (3Q) ground state. This 3Q tetrahedral structure, which critically depends on the Co content, yields the highest-density Skyrmion lattice with scalar spin chirality, resulting in a noticeable anomalous Hall effect. In this work, we demonstrate control of this topological phase via ionic gating. Using four $Co_xTaS_2$ devices with different Co compositions, we show that ionic gating can cover the entire 3Q topological phase and reveal the nature of the thermodynamically inaccessible phase space. Another striking finding in our data is the existence of an adiabatic discontinuity in the phase boundary between the 3Q and 1Q phases. Our work constitutes one of the first examples of electrical control of scalar spin chirality using an antiferromagnetic metal.

Keywords: Topological 3Q tetrahedral state, Scalar spin chirality, Intercalated van der Waals

$Co_{1/3}TaS_2$, Ionic gating, Electrical control, Antiferromagnetic metal.

## Introduction

Magnetism has been crucial in the modern understanding of materials and their Hamiltonian. In particular, two-dimensional (2D) magnetism has been essential, providing arguably the most fundamental understanding: the prime examples include the Ising, XY and Heisenberg models[1-4], all for two dimensions. Despite the immense interest and importance given to 2D magnetism, the experimental progress has been slow in coming. Van der Waals (vdW) magnets, whose monolayer magnetism was first reported in 2016, marked an important step forward[5-7]. With this new class of materials, we can examine the magnetic properties of real materials that exhibit the three 2D Hamiltonians. Meanwhile, vdW magnets also provide extraordinary platforms for next-generation electronics and spintronics, including spin-orbit torque memory and in-memory computing[8-18]. With several important experimental reports, all the experimental tests of 2D magnetism have now been largely completed, leaving only a few remaining questions.

One of the most promising topics that can be investigated using vdW 2D magnets is the 2D topological classes with broken time-reversal symmetry. The first experimental realisation of such is Cr-doped $(Bi,Sb)_2Te_3$[19], where the quantised anomalous Hall effect was found. Another notable example is $MnBi_2Te_4$, whose layer-dependent axion insulator and quantum anomalous Hall ground states have been extensively studied[20-23]. Unlike Cr-doped $(Bi,Sb)_2Te_3$, the stoichiometric nature of $MnBi_2Te_4$ has significantly improved the inhomogeneity issue that can disguise otherwise inherent magnetic and electronic properties. Indeed, although it is difficult, one can grow single-crystal $MnBi_2Te_4$. Experimentally[21-23], it was shown that the antiferromagnetic ordering of $MnBi_2Te_4$ is invariant to the combination of the time-reversal and primitive-lattice translation symmetries, giving rise to a $Z_2$ topological classification. It is essential to note that these so-called magnetic topological insulators rely on the presence of spin-orbit coupling (SOC).

Meanwhile, a completely different route exists to realise the magnetic topology in two dimensions, which takes advantage of topological spin textures. A few representative examples of such are magnetic Skyrmions and vortex crystals[24]. Here, topology occurs through the Berry phase in real space due to the real-space spin winding. These 2D spin configurations are typically observed on hexagonal lattices as a 3Q magnetic state, characterised by a long-range order formed through a linear combination of three distinct spiral configurations[25-31]. Notably, unlike the Chern or axion insulator phases observed in Cr-doped $(Bi,Sb)_2Te_3$ and $MnBi_2Te_4$, which manifest the topological character of these spin textures in macroscopic bulk properties, the presence of SOC is not required.

$Co_{1/3}TaS_2$, an intercalated vdW system with layered magnetic triangular lattices, has recently been suggested as a promising platform for investigating topological physics in two dimensions. A decade ago, theoretical studies predicted the possible formation of a unique 3Q magnetic ground state with a metallic triangular lattice using a ferromagnetic Kondo lattice model[32,33]. This 3Q ground state consists of four spin sublattices that develop a non-coplanar spin configuration equivalent to the geometry of a regular tetrahedron[32]. Notably, they have shown that, for a strong coupling limit with a three-quarters (3/4) filled Fermi surface, the ferromagnetic Kondo lattice model with this 3Q ground state hosts a Chern insulator phase without relativistic spin-orbit coupling (SOC). Remarkably, their prediction of this 3Q ground state has been experimentally reported in stoichiometric bulk $Co_{1/3}TaS_2$[34,35]. Using a combination of various experimental techniques, including neutron scattering, two independent groups have reported that $Co_{1/3}TaS_2$ hosts the four-sublattice 3Q state[34,35]. Importantly, the angle-resolved photoemission spectroscopy (ARPES) result has revealed that the Fermi surface geometry of $Co_{1/3}TaS_2$ is very close to the 3/4-filled Fermi surface, as suggested by theoretical works, highlighting the importance of the Fermi surface in stabilising the observed 3Q state[34].

The topologically nontrivial character of this 3Q tetrahedral state can be understood based on the concept of scalar spin chirality. When one maps the four sublattices of the 3Q state into a single tetrahedron, it becomes clear that this configuration spans a full solid angle of a sphere, representing an all-out state. In the adiabatic limit, this acts as a fictitious flux quantum analogous to that of a magnetic monopole at the centre of the tetrahedron. Applying a time-reversal operation to the all-out state leads to another equivalent state, an all-in state with a reversed sign of the monopole. Thus, this monopole can be viewed as a source of the fictitious magnetic field generated by this 3Q state. The interaction between localised moments and conduction electrons under such a fictitious magnetic field produces an anomalous Hall effect (usually referred to as the topological Hall effect). Notably, this manifestation can occur without relativistic SOC, which marks a clear distinction from the Cr-doped $(Bi,Sb)_2Te_3$ and $MnBi_2Te_4$ systems.

Another interesting finding of $Co_{1/3}TaS_2$ is that the 3Q state is highly sensitive to the cobalt (Co) concentrations. A recent experiment[36] shows that the 3Q state is stable only for Co concentrations less than 1/3, whereas the 3Q state and concomitant anomalous Hall effect disappear for more than 1/3 Co doping. Based on the intricate connection between the Fermi surface and this 3Q ground state, this outcome was understood in terms of the change in the Fermi level resulting from Co doping. As the condition of 3/4 filling of the Fermi surface is bound to change upon Co doping (i.e., adding or removing electrons), Co concentrations beyond 1/3 will result in a larger Fermi surface with a higher conduction carrier density. These results suggest that the electron density can play a very important role in controlling the 3Q magnetism in $Co_{1/3}TaS_2$. While the study of Co-doping dependence opens up an interesting possibility for the electrical control of topological states, whether the influence of adding magnetic vacancies or impurities will result in Fermi surface-level tuning is unclear. Notably, interstitial Co atoms for $x > 1/3$ should intercalate into the new Wyckoff sites in $Co_xTaS_2$, which can modify its character to an unexpectedly large extent beyond the Fermi level shift. Thus, a more controlled experiment with the electron density, without significantly distorting the structure of the stoichiometric case ($x = 1/3$), is required to understand this interesting physics better. Electrical ionic gating is the most promising method for systematically controlling the carrier density, thereby possessing the potential to switch or even control the 3Q topological state. Notably, this technique[37-42] requires fabricating nm-thick devices of a material of interest[40-42], which can be achieved in $Co_xTaS_2$ because of its layered nature[43].

In this study, we have explored this idea by conducting extensive gating experiments on four representative types of $Co_xTaS_2$ samples with x values of 0.299, 0.319, 0.327 and 0.33. All three samples, with smaller x values of 0.299, 0.319, and 0.327, exhibit the topological 3Q magnetic ground state in their pristine forms as bulk single crystals, while the $x = 0.33$ sample sits deep in the 1Q phase. However, as they are located at different positions in the phase diagram, $x = 0.299$ is the lowest co-doping sample with the thermodynamically stable 3Q state, while $x = 0.319$ is at the centre of the 3Q state. At the same time, $x = 0.327$ is closest to the phase boundary, facing a miscibility gap between the topologically nontrivial 3Q state and the topologically trivial spiral state for larger Co doping[36]. Note that $x = 0.33$ is located in the 1Q state. By gating, we have shown that one can switch on and off the 3Q state and control its topological phases in the 3Q phase. These findings provide new insights and opportunities in both fundamental physics problems and spintronic/electronic device applications.

## Results and Discussion

### Scalar spin chirality and anomalous Hall effect in $Co_{1/3}TaS_2$ nanodevices

As illustrated in Fig. 1a, $Co_{1/3}TaS_2$ exhibits a hexagonal structure with a space group of $P6_322$, where the Co atoms are intercalated into the vdW gap of the 2H-$TaS_2$[34,36]. The intercalated Co atoms induce magnetism into this layered material and form two antiferromagnetic long-range orders at two different temperature ranges, separated by two phase transitions at $T_{N1}$ = 38 K and $T_{N2}$ = 26.5 K[34,35,43]. Figure 1b shows the schematic for the spin configuration of the antiferromagnetic ground state for $T < T_{N2}$[34,36]. Interestingly, this ground state of $Co_{1/3}TaS_2$ possesses a 3Q spin texture, featuring a tetrahedral structure of four "all-out" spins standing on its vertex (Fig. 1c). As described in the Introduction section, the tetrahedral spin units carry topological properties characterised by scalar spin chirality aligned ferromagnetically across the entire 3D structure, according to recent neutron diffraction studies[34-36]. The key essence is that the three non-coplanar spins on a single triangular plaquette produce a gauge flux characterised by the scalar spin chirality (Fig. 1d). Each triangular plaquette generates a local gauge flux due to the non-collinear arrangement of spins, with the sign of $\chi_{ijk}$ indicated as $+$ or $-$. These local contributions do not cancel out because the scalar spin chirality is uniformly aligned across the entire lattice, leading to a finite net gauge flux. When an electron (or hole) carrier moves around these three non-coplanar spins, it acquires an additional phase in its wave function, called a Berry phase, in real space. Such scalar spin chirality triggers a transverse motion of carriers without requiring an external magnetic field, consequently leading to a spin-texture-governed anomalous Hall effect. The tetrahedral spin configuration, where each spin points outward (or inward) from the center, ensures that the gauge flux is preserved across the lattice, contributing consistently to the AHE. As depicted in Fig. 1e, the scalar spin chirality can reverse its sign under an external magnetic field when the sweeping magnetic field alters the spin pointing directions oppositely, leading to a significant anomalous Hall loop for $Co_{1/3}TaS_2$[34,36].

$Co_{1/3}TaS_2$ nanoflakes, approximately 20 nm in thickness (Fig. S1), were exfoliated from high-quality single crystals and used to fabricate nanodevices for transport measurements (see Methods for more details). Figure 1f shows the typical anomalous Hall measurement results of the pristine $Co_{0.299}TaS_2$ nanodevices, i.e. the transverse Hall resistivity $\rho_{xy}$ as a function of a magnetic field at various temperatures. A prominent anomalous Hall loop develops at 7 K and gradually suppresses as the temperature increases up to 20 K. As indicated in the phase diagram of Fig. 2e, the transition from the 1Q phase to the 3Q phase occurs at approximately 20 K. Below this temperature, the system stabilizes into a 3Q spin configuration, which generates scalar spin chirality and gives rise to the AHE. This transition is reflected in Fig. 1f, where a prominent hysteresis loop in the anomalous Hall resistance appears at temperatures below 20 K, indicating the influence of the 3Q state on the AHE.

Additionally, the phase diagram in Fig. 2e shows that above 40 K, the system enters a paramagnetic state, where magnetic ordering vanishes and the AHE should disappear. Indeed, in Fig. 1f, the AHE signal gradually diminishes with increasing temperature and nearly vanishes between 25 and 40 K, consistent with the loss of magnetic ordering in the paramagnetic phase. Therefore, the temperature-dependent behavior of the AHE in Fig. 1f aligns well with the phase transitions described in Fig. 2e. The anomalous Hall resistivity $\rho_{AHE}$ at zero magnetic field and the coercive field $H_c$ at half switching are extracted from Fig. 1f and plotted as a function of temperature in Fig. 1 g-h. Both show a gradual reduction from a large value to nearly zero at around 25 K with increasing temperature, highlighting a magnetic transition behaviour that reflects the absence of scalar spin chirality in the 1Q phase. These observations qualitatively agree with the bulk transport measurement result, indicating that the 20 nm-thick $Co_{1/3}TaS_2$ sample has the same 3Q magnetic ground state as that reported in bulk studies[34,35,43].

**Ionic gating on $Co_{1/3}TaS_2$ nanoflakes**

We present the manipulation of the topological 3Q structure in $Co_{1/3}TaS_2$ nanoflakes by employing the ionic gating technique[40-42]. The key idea is that since the anomalous Hall effect is the direct consequence of the underlying 3Q spin texture, measuring this quantity under gated voltage allows tracking of how this magnetic order changes under ionic gating. Figure 2a shows the 3D cartoon schematic of the gating device with a Hall electrode geometry. The ionic gel-like electrolyte is dropped onto the device, and the gate voltage $V_G$ is applied between the neighbouring gating pad (G) and the drain end (D). The current flows between the source end (S) and the drain end (D) while the transverse Hall voltage is monitored simultaneously. Figure 2b represents the optical image of a typical real gating device, where the gate window partially covers the $Co_{1/3}TaS_2$ nanoflake and the gate pad (G) is physically disconnected from the nanoflake. One can understand the ionic gating effect in Fig. 2c-d: positive gate voltage drives the $Li^+$ ion to accumulate at the $Co_{1/3}TaS_2$ nanoflake, and then the $Li^+$ ion diffuses into the gap between each vdW layer. These $Li^+$ ions attract and induce more electrons into the system, leading to electron doping. On the other hand, a negative gate voltage will push and accumulate $ClO4^-$ ions on the nanoflake, deplete electrons, and induce more hole carriers to diffuse into the device, causing hole doping. Although the above two processes are not exact reversals of each other in terms of carrier doping efficiency, qualitative electron and hole doping can be expected, respectively.

Figure 2e illustrates the magnetic phase diagram of $Co_{1/3}TaS_2$, where the paramagnetic, 1Q, 3Q, and helical phases dominate different regions of the phase space in terms of temperature and Co concentration. Here, we focus on the 3Q phase regime that spans the Co concentration range below 1/3. As the Co concentration changes from 0.29 to 0.327, the 3Q ground state is initially strengthened due to a decrease in vacancies but subsequently weakened, entering a competing phase as it stabilises an entirely different helical order for $x > 1/3$[34]. Although real situations can be much more complicated, a simple view is that increasing the number of Co atoms generally leads to an increased conduction electron density (intercalated Co atoms become divalent $Co^{2+}$ ions, thereby providing two electrons to the system). In the gating experiment, we apply a gate voltage $V_G$ to induce electron or hole doping into $Co_{1/3}TaS_2$ and investigate the anomalous Hall response. We prepared four samples with different Co concentrations of 0.299, 0.319, 0.327, and 0.33, and applied either a positive or negative gate voltage or both. For the $Co_{0.299}TaS_2$ close to the left boundary of the 3Q regime, we anticipate that positive gate voltage will strengthen the 3Q phase by electron doping. On the other hand, negative gate voltage weakens the 3Q state by hole doping. For the $Co_{0.319}TaS_2$ in the middle of the 3Q regime, the 3Q phase remains stable and thus would not be much affected by electrical gating. On the other hand, positive gating-induced electron doping is expected to push the 3Q phase of $Co_{0.327}TaS_2$ to the right side of the 3Q regime. Finally, we test the gating effect on $Co_{0.33}TaS_2$, which is already in the trivial helical phase.

Figure 3 exhibits the electrical response under gating of the $Co_{0.299}TaS_2$ nanodevices. As can be seen in Fig. 3a-b, the longitudinal resistance $R_{xx}$ can be effectively tuned by the ionic gating, indicating the induced electron doping at the positive gate and hole doping at the negative gate, which is consistent with the analysis illustrated in Fig. 2. We attribute the increase in resistance under negative gate voltage to a reduction in carrier density and mobility by ionic gating. Strikingly, the anomalous Hall loops have been substantially modulated under the ionic gating (Fig. 3c). At $V_G$ = 0 V, the $\rho_{xy}$-magnetic field curve shows a clear anomalous Hall loop, consistent with the measurement result of $Co_{0.299}TaS_2$ nanoflake in Fig. 1f. The anomalous Hall effect is enhanced by applying the positive gate voltages and suppressed by the negative ones. The anomalous Hall resistivity, $\rho_{AHE}$, is further extracted as a function of gate voltage for three different devices with the same Co concentration of 0.299. Interestingly, they collapse into a universal evolution trend, such that the anomalous Hall resistivity $\rho_{AHE}$ increases at positive gating but reduces to zero at negative gating. This observation is in good agreement with our expectation, as illustrated in Fig. 2e. In the inset, we show the effect of gating on the Fermi surface of $Co_{1/3}TaS_2$, particularly from the perspective of its nesting mechanism that intimately relates to the 3Q state. $Co_{0.299}TaS_2$ is expected to have the Fermi surface relatively shrunk compared to the nearly 3/4-filling in the stoichiometric sample. Under negative $V_G$, the Fermi surface filling will further deviate from its ideal nesting filling with the M-point Fermi wavevector, eventually destroying the 3Q ground state. This observation implies that the system under negative voltage may transition either to a paramagnetic state or to another magnetically ordered state that does not host scalar spin chirality, including the 1Q state. Intuitively, however, it is most likely to be a magnetically ordered 1Q state, given that our experiment was conducted at a low temperature of 7 K. Additionally, there is a 1Q state that separates 3Q and paramagnetic states for temperatures between 38 and 20 K in pristine samples. Similarly, the coercive field $H_c$ (Fig. 3e) also retains significant values against the gate voltages where the anomalous Hall effect maintains its presence, but vanishes at gate voltages where the anomalous Hall effect disappears.

Figure 4 summarises the two other experimental results of gating measurements with $Co_{0.319}TaS_2$ and $Co_{0.327}TaS_2$ nanoflakes. As shown in Fig. 4a, the anomalous Hall loops remain prominent regardless of the applied gate voltages, although subtle changes occur due to ionic gating. The extracted anomalous Hall resistivity $\rho_{AHE}$ and coercive field $H_c$ also remain significant during the gating experiments. This strong robustness of the anomalous Hall effect against gating agrees well with the expected behaviour of the 3Q ground state in $Co_{0.319}TaS_2$, based on the magnetic phase diagram of Fig. 2e. The $Co_{0.327}TaS_2$ sample located close to the competing phases region, i.e., a phase boundary between the 3Q and the helical phases in Fig. 2e, shows behaviour similar to that of $Co_{0.319}TaS_2$ under gate voltages, where the pristine $Co_{0.327}TaS_2$ nanoflake exhibited a larger coercive field compared to samples with smaller x, which trend is consistent with findings from a previous study on bulk samples[36]. In our experiment, we observed that the anomalous Hall loops remained stable under positive gate voltages, as shown in Fig. 4d. Both the anomalous Hall resistivity $\rho_{AHE}$ and the coercive field $H_c$ remained unchanged in magnitude with increasing positive gate voltages, as presented in Fig. 4e-f. Although positive gating shifts the carrier density away from the 3/4-filling suggested to be crucial for stabilizing the 3Q state in $Co_{1/3}TaS_2$, the overdoped $Co_{0.34}TaS_2$ sample exhibits a significantly different Fermi surface shape with Fermi wavevector toward the K-point compared to the underdoped $Co_{0.325}TaS_2$ sample, which retains a stable 3Q phase, based on our recent ARPES measurements. These results indicate that the Fermi surface in the overdoped regime undergoes an adiabatic discontinuity that cannot be reached by sample carrier doping via positive gate voltage. A passing note: the Fermi surface changes induced by the gating can generally be expected to affect the Ta 5d band.

Generally, gating (i.e., carrier doping) directly causes a shift in the Fermi level, thereby naturally modifying the Fermi surface. Such a change in the Fermi surface can correspondingly affect the topological spin texture in $Co_xTaS_2$ since its topological 3Q states depend on proximity to the 3/4 filling of the Fermi surface. Simultaneously, the Fermi level shift will unavoidably change the Berry curvature integration over occupied bands throughout the Brillouin zone. These two effects are not easily separable but are rather deeply linked in general. Specifically, the 3Q tetrahedral spin texture that underlies the intrinsic AHE is known to be highly sensitive to the Fermi surface geometry, as evidenced by both theoretical proposals[32,34] and the Co composition dependence[36]. Therefore, tuning the carrier density via gating can lead to a modification of the spin texture itself, which directly impacts the Hall signal. Simultaneously, the Berry curvature—determined by the electronic bands below the Fermi level over the whole Brillouin zone—would also change correspondingly, altering the magnitude of the AHE. We believe that the observed gating-induced changes in AHE are best interpreted as the results of both effects acting together. That is, gating not only shifts the Fermi level (affecting the Berry curvature) but also perturbs the delicate stability of the 3Q order (affecting the presence of topological spin texture). This viewpoint aligns with the motivation behind our gating experiments, which aimed to explore the interplay between carrier density, magnetic order, and topological transport. As a passing remark, AHE remains absent during the gating process on $Co_{0.33}TaS_{2}$, as it already resides deep in the trivial helical phase.

In summary, we have reported the electrical control of the topologically nontrivial 3Q tetrahedral state in $Co_xTaS_2$ with $x$ =0.299, 0.319, and 0.327 by varying the conduction electron density through gating voltage. This work demonstrates the full electrical control of the topological 3Q magnetic structure observed in optimally doped $Co_{1/3}TaS_2$, which emerges without the relativistic spin-orbit coupling (SOC) term. The gating-induced change in the transverse Hall voltage indicates that electron density is directly responsible for the modified magnetic ground states. When the systems remain in the 3Q state, all samples have a strong anomalous Hall effect. On the other hand, when the electron density is significantly displaced from the stable 3Q state, this anomalous Hall effect is markedly suppressed. This demonstration agrees with the suggestion that the key factor for a 3Q tetrahedral ground state is the Fermi surface geometry of $Co_{1/3}TaS_2$, whose 3/4-filling can induce a nesting effect with the M-point Fermi wavevector. This work is a unique demonstration of how one can control the nonrelativistic SOC-free topological magnetic order with scalar spin chirality in a metallic antiferromagnet by modulating the carrier density.

## Methods

### Sample preparation and characterisation

High-quality single-crystal $Co_{1/3}TaS_2$ is prepared by the two-step chemical vapour transport (CVT) method. Polycrystalline $Co_{1/3}TaS_2$ is prepared using the following process: A mixture of Co (Alfa Aesar, > 99.99%), Ta (Sigma Aldrich, > 99.99%), and S (Sigma Aldrich, > 99.999%) in a molar ratio of $1\pm x$:3:6. Composition used for the experiment could be obtained when additional Co added in the first step around $x$ = 1.5 – 2.5 %. Homogeneous mixed powders were sealed with a quartz tube of 13 cm in length with an inner (outer) diameter of 17 (20) mm. Argon gas was charged to around 2 torr in the quartz tube. The quartz tube was heated at a rate of 100 °C/hr until it reached 900 °C and then maintained at this temperature for 5 days. After 5 days, the furnace was turned off and allowed to cool naturally.

Fabricated polycrystals are ground and used for CVT reactions under the same conditions as quartz tubes. As a transport agent, 200 mg of $I_2$ was added with 2.2 g of polycrystalline $Co_{1/3}TaS_2$. The quartz tube was heated at a rate of 100 °C/hr, and the hot (cold) zone temperature was maintained at 940 (860) °C for 2 weeks. After 2 weeks, the furnace was turned off and allowed to cool naturally. A shiny hexagonal-shaped single crystal is obtained, and each crystal was characterised by a superconducting quantum interference device magnetometer, MPMS-XL5 (Quantum Design, USA), along with electrical transport measurements using the standard four-probe method with CFMS (Cryogenics Ltd, UK).

### Gating experiments

The fabrication of the gating device began with the standard process applied to an exfoliated $Co_{1/3}TaS_2$ nanoflake. A large pad for the Hall electrodes was patterned to facilitate wire bonding later. An additional gate pad, unconnected to the target sample, was also included to apply a gate voltage. To prevent leakage current between the gate electrode and unwanted areas, an insulating PMMA layer was spin-coated over the patterned metal electrodes. A second pattern was then defined in the gate window regions using lithography, allowing the gate voltage to affect only a specific area of the nanoflake while avoiding the Hall measurement electrodes.

Next, a lithium-based solid electrolyte ($LiClO_4$/PEO/methanol) was prepared by dissolving 0.3 g of $LiClO_4$ and 1 g of PEO (Mw = 100,000) in 15 ml of methanol, then stirring overnight at 50 °C[40]. The electrolyte was applied to the substrate, which was subsequently annealed at 95 °C inside the glove box for at least one hour to evaporate the methanol and solidify the electrolyte.

A gate voltage ($V_G$) was applied using a Keithley 2400 source meter, swept at fixed temperatures between 320 and 360 K in a vacuum inside a cryostat. The sample was kept at these temperatures for at least 30 minutes to stabilise the gating effect. The resistance was measured in both longitudinal and transverse directions using a lock-in amplifier (Stanford Research SR830). Afterwards, the temperature was lowered to ~4 K and warmed to 7 K for anomalous Hall measurements under gating. The magnetic field was applied along the out-of-plane direction.

**Data Availability**

The data that support the findings of this study are available from the corresponding author upon reasonable request.

**Acknowledgements**

We acknowledge Suhan Son, Hyeonsik Cheong, Beom Hyun Kim, Jaehoon Kim, Young-Woo Son, Moon-Sun Nam, Arzhang Ardavan, and Cristian Batista for their helpful discussions. We also thank Ding Zhang, Ke He, and Qi-Kun Xue for sharing their experimental knowledge and generous help at some stage of this work. We are particularly indebted to Hyeonsik Cheong for the critical manuscript reading and valuable comments. This work was supported by the Samsung Science & Technology Foundation (Grant No. SSTF-BA2101-05). One of the authors (J.-G.P.) is partly funded by the Leading Researcher Program of the National Research Foundation of Korea (Grant No. RS-2020-NR049405). P.P. acknowledges support by the U.S. Department of Energy, Office of Science, Basic Energy Sciences, Materials Science and Engineering Division.

**Author contributions**

J.-G.P. initiated and supervised the project. P.P. and WHC synthesised the single-crystal samples and performed all the bulk characterisations. JHK and HCK conducted the gating experiments and analysed the data with KZ and HJN. JHK, PP, KZ, and J.-G.P. wrote the manuscript with contributions from all authors.

**Competing interests**

The authors declare no competing interests.

## Figures

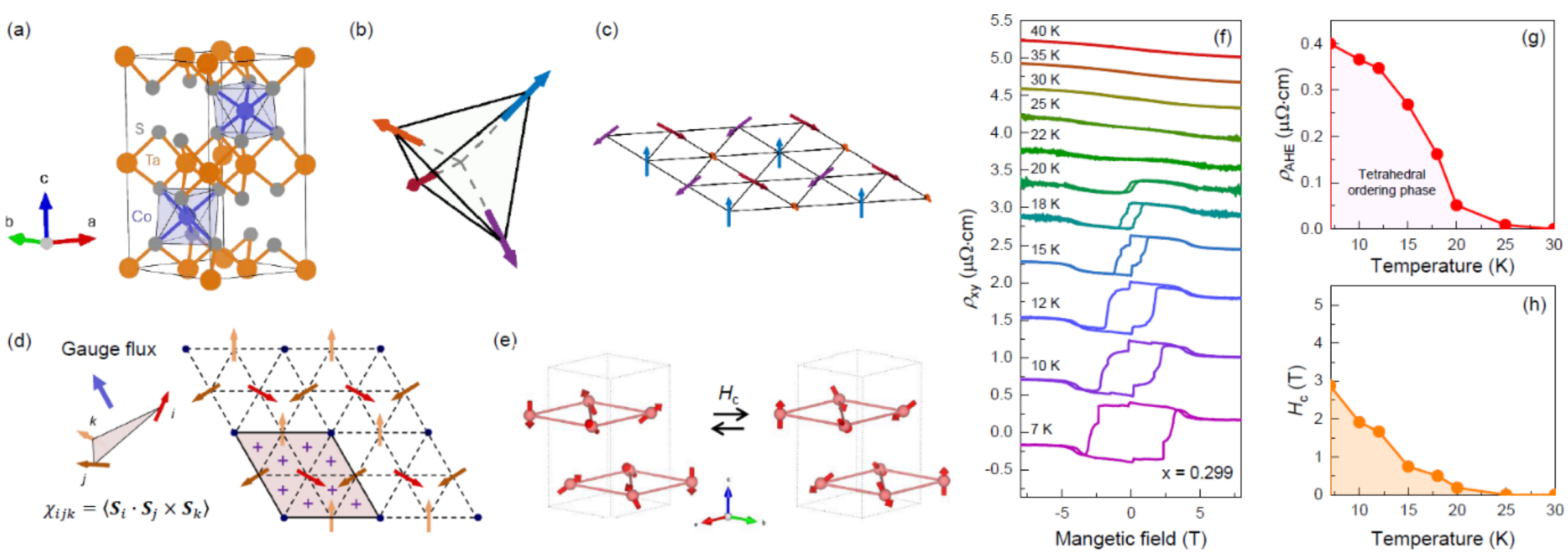

**Fig 1 | Illustration of topological scalar spin chirality and anomalous Hall effect in pristine $Co_{1/3}TaS_2$ nanoflakes. a**, Crystal structure of $Co_{1/3}TaS_2$. **b**, Tetrahedron formed by four spin sublattices. **c**, Triangular lattice of Co spins in the 2D plane. **d**, Effective gauge flux formed by three non-coplanar spins, corresponding to the scalar spin chirality defined as $\chi_{ijk} = \langle \boldsymbol{S_i} \cdot (\boldsymbol{S_j} \times \boldsymbol{S_k}) \rangle$, where $\boldsymbol{S_i}$, $\boldsymbol{S_j}$ and $\boldsymbol{S_k}$ represent spins at the vertices of a triangular plaquette. The sign of $\chi_{ijk}$ (indicated as $\pm$) is determined by the spin configuration, with the triangular plaquettes colored to indicate their corresponding chirality. **e**, Tetrahedral spin configuration, having opposite scalar spin chirality to each other, can be switched by an external magnetic field at the coercive field $H_c$. **f**, Magnetic field dependence of Hall resistivity at various temperatures for a $Co_{0.299}TaS_2$ nanoflake. **g**, Temperature dependence of anomalous Hall resistivity at zero magnetic field. **h**, Temperature dependence of coercive field $H_c$ at half switching.

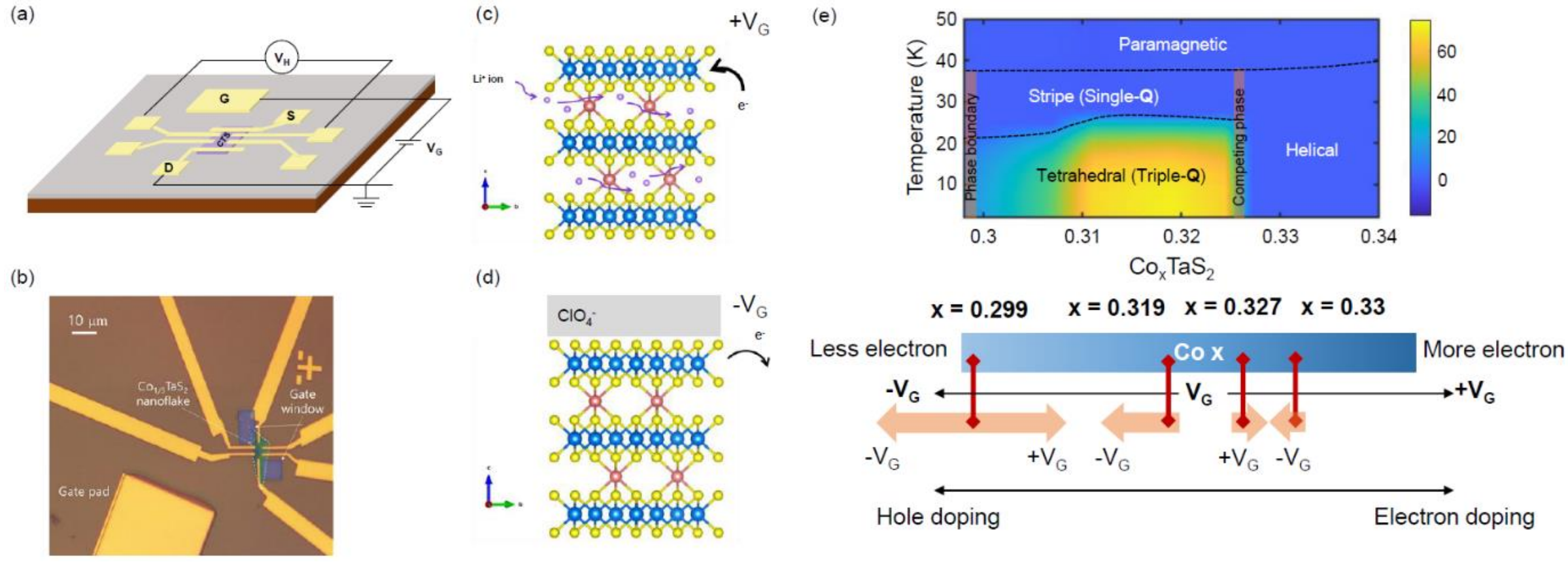

**Fig. 2 | Ionic gating illustration on $Co_{1/3}TaS_2$ nanoflakes and manipulation mechanism of quantum 3Q phase. a**, Hall measurement schematic of a gating device, while the gate voltage was applied from the side gate pad to the drain. **b**, Optical image of a typical gating device of $Co_{1/3}TaS_2$ nanoflake. **c-d**, Gating mechanism illustration of ionic gating. A positive gate voltage causes the intercalation of lithium ions passing through the material, inducing electron doping. In contrast, the negative voltage accumulates the $ClO_4^-$ ions, depleting electrons and introducing hole doping, which diffuses throughout the system. **e**, Phase diagram of $Co_{1/3}TaS_2$ (adapted from a very recent work[36]) and corresponding schematic of electron/hole doping. The colour plot of the Hall conductivity in $Co_xTaS_2$ bulk crystals constructs the magnetic phase diagram, which comprises four distinct phases: Paramagnetic, Single-Q (1Q), Triple-Q (3Q), and Helical phases. The anomalous Hall effect originates from the scalar spin chirality in 3Q phase states and varies with Co concentration. The red pointers indicate four kinds of gating samples with different Co concentrations. In our experiments, the orange arrows schematically indicate the gating regions by positive and negative voltages.

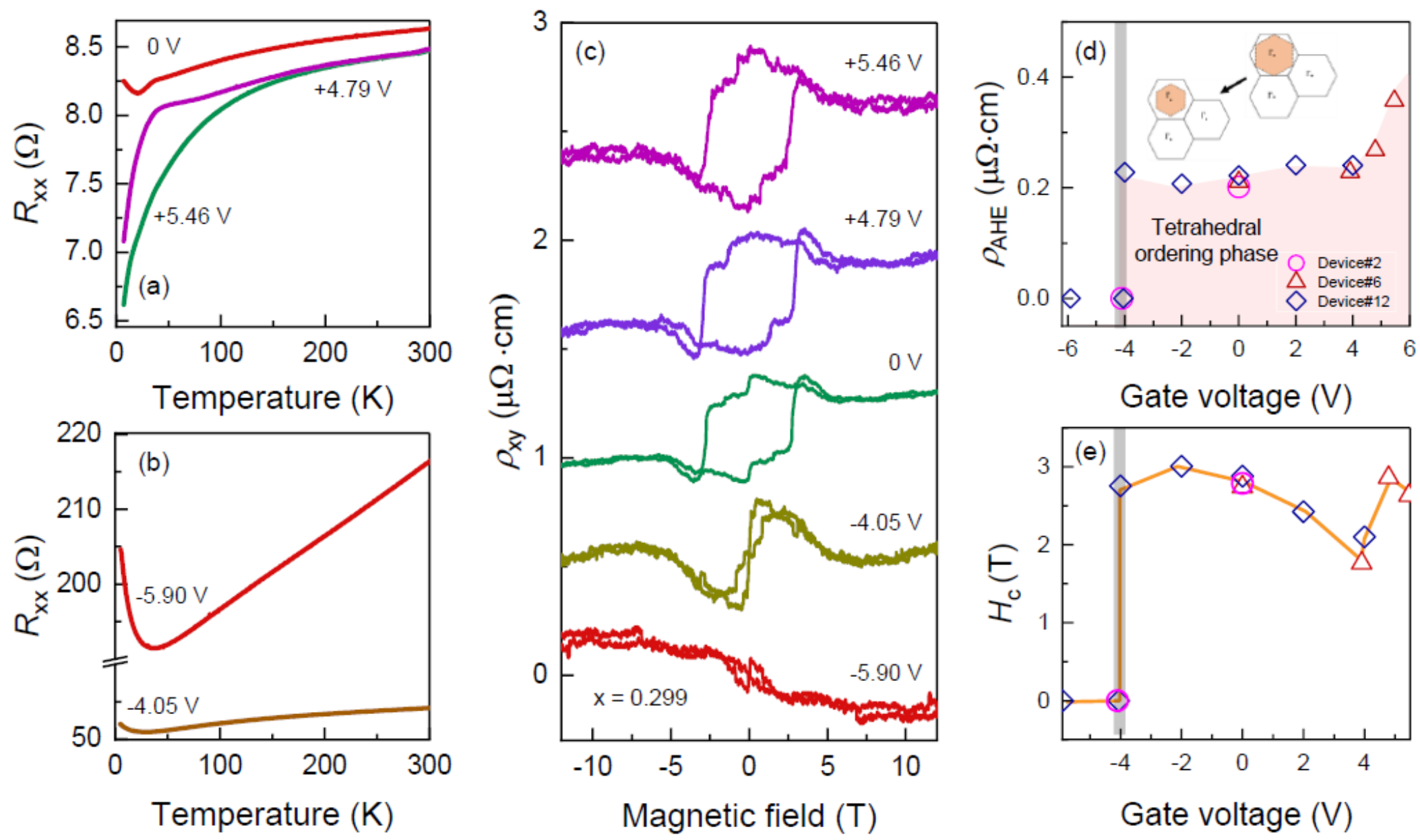

**Fig 3 | Gating responses of $Co_{0.299}TaS_2$ nanoflakes. a-b**, Longitudinal resistance $R_{xx}$ as a function of temperature under different gate voltages. **c**, Hall resistivity $\rho_{xy}$ versus magnetic field at different gate voltages under 7 K. The magnetic field is applied out-of-plane and swept from -12 to 12 T. **d**, Gate-voltage dependence of the anomalous Hall resistivity $\rho_{AHE}$ at zero magnetic field. The 3Q tetrahedral ordering exists in the pink-coloured region where the value of the anomalous Hall resistivity is finite. The inset depicts the Fermi surface filling modulated by gate voltage. **e**, Gate-voltage dependence of the coercive field $H_c$. The grey line indicates where the tetrahedral ordering disappears.

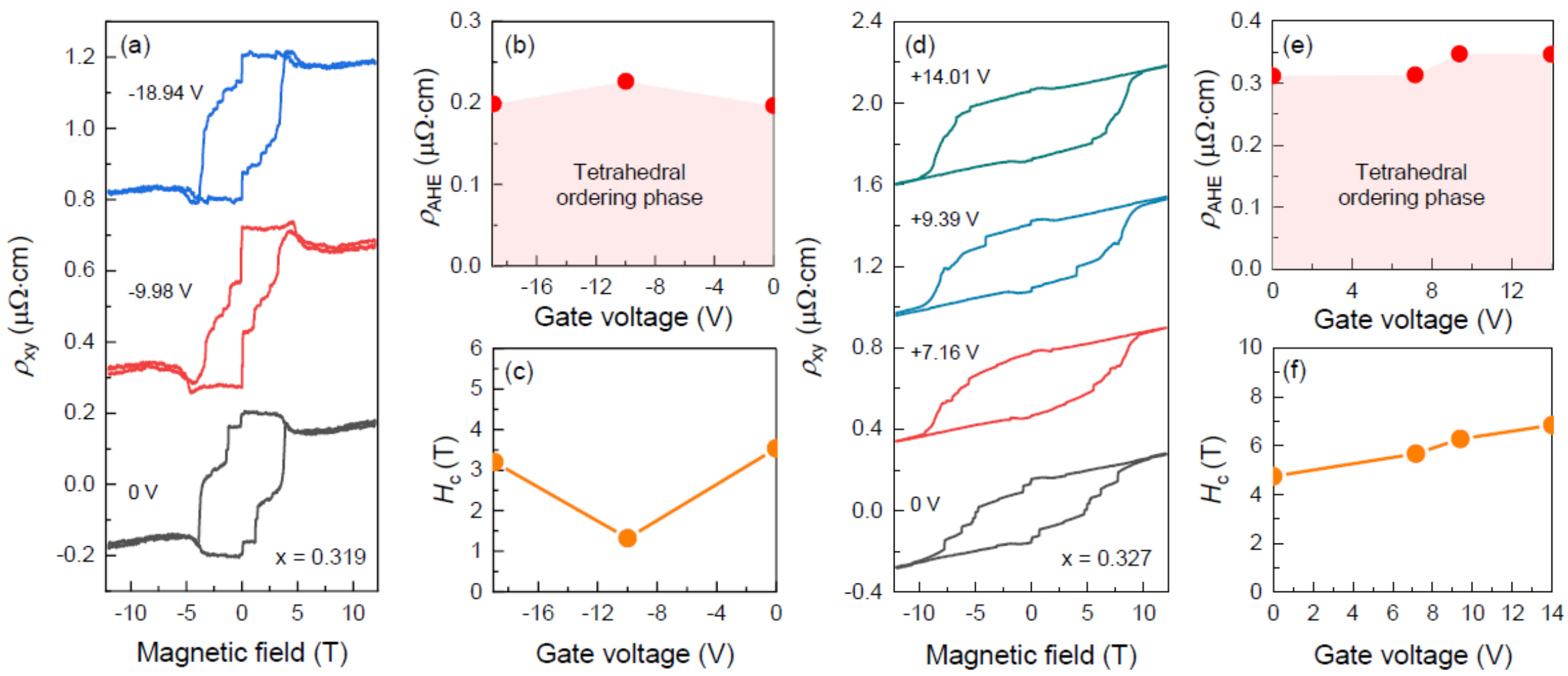

**Fig 4 | Gating responses of $Co_{0.319}TaS_2$ and $Co_{0.327}TaS_2$ nanoflakes. a,d** Hall resistivity $\rho_{xy}$ versus magnetic field at different gate voltages for $Co_{0.319}TaS_2$ (**a**) and $Co_{0.327}TaS_2$ (**d**) under 7 K, respectively. The magnetic field is applied out-of-plane. **b,e** Gate-voltage dependence of the anomalous Hall resistivity $\rho_{AHE}$ at zero magnetic field for $Co_{0.319}TaS_2$ (**b**) and $Co_{0.327}TaS_2$ (**e**), respectively. **c,f** Gate-voltage dependence of the coercive field $H_c$ for $Co_{0.319}TaS_2$ (**c**) and $Co_{0.327}TaS_2$ (**f**), respectively.